\documentstyle[12pt]{article}
\begin{document}

\def\ds{\displaystyle}
\def\beq{\begin{equation}}
\def\eeq{\end{equation}}
\def\bea{\begin{eqnarray}}
\def\eea{\end{eqnarray}}
\def\beeq{\begin{eqnarray}}
\def\eeeq{\end{eqnarray}}
\def\ve{\vert}
\def\vel{\left|}
\def\ver{\right|}
\def\nnb{\nonumber}
\def\ga{\left(}
\def\dr{\right)}
\def\aga{\left\{}
\def\adr{\right\}}
\def\lla{\left<}
\def\rra{\right>}
\def\rar{\rightarrow}
\def\nnb{\nonumber}
\def\la{\langle}
\def\ra{\rangle}
\def\ba{\begin{array}}
\def\ea{\end{array}}
\def\tr{\mbox{Tr}}
\def\ssp{{\Sigma^{*+}}}
\def\sso{{\Sigma^{*0}}}
\def\ssm{{\Sigma^{*-}}}
\def\xis0{{\Xi^{*0}}}
\def\xism{{\Xi^{*-}}}
\def\qs{\la \bar s s \ra}
\def\qu{\la \bar u u \ra}
\def\qd{\la \bar d d \ra}
\def\qq{\la \bar q q \ra}
\def\gGgG{\la g^2 G^2 \ra}
\def\q{\gamma_5 \not\!q}
\def\x{\gamma_5 \not\!x}
\def\g5{\gamma_5}
\def\sb{S_Q^{cf}}
\def\sd{S_d^{be}}
\def\su{S_u^{ad}}
\def\rl{\hat{m}_{\ell}}
\def\ss{\hat{s}}
\def\rr{\hat{r}_{K_1}}
\def\sbp{{S}_Q^{'cf}}
\def\sdp{{S}_d^{'be}}
\def\sup{{S}_u^{'ad}}
\def\ssp{{S}_s^{'??}}
\def\sig{\sigma_{\mu \nu} \gamma_5 p^\mu q^\nu}
\def\fo{f_0(\frac{s_0}{M^2})}
\def\ffi{f_1(\frac{s_0}{M^2})}
\def\fii{f_2(\frac{s_0}{M^2})}
\def\O{{\cal O}}
\def\sl{{\Sigma^0 \Lambda}}
\def\es{\!\!\! &=& \!\!\!}
\def\ar{&+& \!\!\!}
\def\ek{&-& \!\!\!}
\def\cp{&\times& \!\!\!}
\def\fhs{Re[FH^*]}
\def\ghs{Re[GH^*]}
\def\bcs{Re[BC^*]}
\def\fgs{Re[FG^*]}
\def\hh{|H|^2}
\def\cc{|C|^2}
\def\gg{|G|^2}
\def\ff{|F|^2}
\def\bb{|B|^2}
\def\aa{|A|^2}
\def\hh{|H|^2}
\def\ee{|E|^2}
\def\rl{\hat{m}_{\ell}}
\def\r{\hat{m}_K}
\def\s{\hat{s}}
\def\ll{\Lambda}

\renewcommand{\textfraction}{0.2}    
\renewcommand{\topfraction}{0.8}

\renewcommand{\bottomfraction}{0.4}
\renewcommand{\floatpagefraction}{0.8}
\newcommand\mysection{\setcounter{equation}{0}\section}
\newcommand{\bra}[1]{\langle {#1}}
\newcommand{\ket}[1]{{#1} \rangle}
\newcommand{\ebar}{{\bar{e}}}
\newcommand{\sbar}{\bar{s}}
\newcommand{\cbar}{\bar{c}}
\newcommand{\bbar}{\bar{b}}
\newcommand{\qbar}{\bar{q}}
\renewcommand{\l}{\ell}
\newcommand{\lbar}{\bar{\ell}}
\newcommand{\psibar}{\bar{\psi}}
\newcommand{\barB}{\overline{B}}
\newcommand{\barK}{\overline{K}}
\newcommand{\thetaK}{\theta_{K_1}}
\newcommand{\onepone}{{1^1P_1}}
\newcommand{\sanpone}{{1^3P_1}}
\newcommand{\kone}{{K_1}}
\newcommand{\barkone}{{\overline{K}_1}}
\renewcommand{\Re}{\mathop{\mbox{Re}}}
\renewcommand{\Im}{\mathop{\mbox{Im}}}
\newcommand{\T}{{\cal T}}
\newcommand{\eff}{{\rm eff}}
\newcommand{\A}{{\cal A}}
\newcommand{\B}{{\cal B}}
\newcommand{\C}{{\cal C}}
\newcommand{\D}{{\cal D}}
\newcommand{\E}{{\cal E}}
\newcommand{\F}{{\cal F}}
\newcommand{\G}{{\cal G}}
\renewcommand{\H}{{\cal H}}
\newcommand{\hats}{\hat{s}}
\newcommand{\hatp}{\hat{p}}
\newcommand{\hatq}{\hat{q}}
\newcommand{\hatm}{\hat{m}}
\newcommand{\hatu}{\hat{u}}
\newcommand{\alphaem}{\alpha_{\rm em}}
\newcommand{\konel}{K_1(1270)}
\newcommand{\koneh}{K_1(1400)}
\newcommand{\barkonel}{\barK_1(1270)}
\newcommand{\barkoneh}{\barK_1(1400)}
\newcommand{\konea}{K_{1A}}
\newcommand{\koneb}{K_{1B}}
\newcommand{\barkonea}{\barK_{1A}}
\newcommand{\barkoneb}{\barK_{1B}}
\newcommand{\mkone}{m_{\kone}}
\newcommand{\konep}{K_1^+}
\newcommand{\konem}{K_1^-}
\newcommand{\konelm}{K_1^-(1270)}
\newcommand{\konehm}{K_1^-(1400)}
\newcommand{\konelp}{K_1^+(1270)}
\newcommand{\konehp}{K_1^+(1400)}
\newcommand{\konelz}{\overline{K}{}^0_1(1270)}
\newcommand{\konehz}{\overline{K}{}^0_1(1400)}
\newcommand{\Bm}{B^-}
\newcommand{\Bz}{\overline{B}{}^0}
\newcommand{\Kstar}{K^*(892)}
\newcommand{\BABAR}{BABAR}
\newcommand{\BELLE}{Belle}
\newcommand{\CLEO}{CLEO}
\newcommand{\leftu}{\gamma^\mu L}
\newcommand{\leftd}{\gamma_\mu L}
\newcommand{\rightu}{\gamma^\mu R}
\newcommand{\rightd}{\gamma_\mu R}
\newcommand{\Br}{{\cal B}}
\newcommand{\sect}[1]{Sec.~\ref{#1}}
\newcommand{\eqref}[1]{(\ref{#1})}
\newcommand{\fig}{FIG.~}
\newcommand{\figs}{FIGs.~}
\newcommand{\tbl}{TABLE~}
\newcommand{\tbls}{TABLEs~}
\newcommand{\errpm}[3]{#1^{+{#2}}_{-{#3}}}
\newcommand{\errpmf}[5]{{#1}^{ +{#2} +{#4} }_{-{#3}-{#5}}}
\newcommand{\lpm}{\l^+\l^-}
\newcommand{\epm}{e^+e^-}
\newcommand{\mupm}{\mu^+\mu^-}
\newcommand{\taupm}{\tau^+\tau^-}
\newcommand{\AFB}{A_{\rm FB}}
\newcommand{\barAFB}{\overline{A}_{\rm FB}}
\newcommand{\GeV}{{\,\mbox{GeV}}}
\newcommand{\MeV}{{\,\mbox{MeV}}}
\newcommand{\degree}{^\circ}
\newcommand{\mB}{m_B}
\newcommand{\SM}{{\rm SM}}
\newcommand{\NP}{{\rm NP}}
\newcommand{\barc}{\bar{c}}
\newcommand{\xipara}{\xi_\parallel^{\kone}}
\newcommand{\xiperp}{\xi_\perp^{\kone}}
\newcommand{\xiparal}{\xi_\parallel^{\konel}}
\newcommand{\xiperpl}{\xi_\perp^{\konel}}
\newcommand{\xiparah}{\xi_\parallel^{\koneh}}
\newcommand{\xiperph}{\xi_\perp^{\koneh}}
\newcommand{\para}{\parallel}
\newcommand{\alphas}{\alpha_s}
\newcommand{\pA}{p_{\kone}}
\newcommand{\lcaption}[2]{\caption{(label:{#2}) #1}\label{#2}}
\newcommand{\Rmunr}{R_{\mu,\rm nr}}
\newcommand{\RdGamma}{R_{d\Gamma/ds,\mu}}
\providecommand{\dfrac}[2]{\frac{\displaystyle
{#1}}{\displaystyle{#2}}}
\def\baeq{\begin{appeq}}     \def\eaeq{\end{appeq}}
\def\baeeq{\begin{appeeq}}   \def\eaeeq{\end{appeeq}}
\newenvironment{appeq}{\beq}{\eeq}
\newenvironment{appeeq}{\beeq}{\eeeq}
%
%
%
%
\title{\boldmath\bf
Lepton polarization in $B \to K_1 \ell^+ \ell^-$ Decays}
\author{\\
{\small V. Bashiry$^1$\thanks {e-mail: bashiry@ciu.edu.tr}}, \\
 {\small $^1$Engineering Faculty,
Cyprus International University,}
\\ {\small Via Mersin 10, Turkey }}
\date{}
\begin{titlepage}
\maketitle \thispagestyle{empty}
\begin{abstract}
We study the single and double lepton polarization asymmetries in
the semileptonic $B$ meson decays $B \to K_1 (1270) \ell^+ \ell^-$
$\ell \equiv e$, $\mu$, $\tau$), where the strange $P$-wave meson,
$K_1(1270)$, is  the mixtures of the $K_{1A}$ and $K_{1B}$, which
are the $1^3P_1$ and $1^1P_1$ states, respectively. The lepton
polarization asymmetries show  relatively strong dependency in the
various region of dileptonic invariant mass.  The lepton
polarization asymmetries can also be used for determining the
$K_1(1270)$--$K_1(1400)$ mixing angle, $\theta_{K_1}$ and new
physics effects.  Furthermore, it is shown that these asymmetries in
$B\to K_1(1270)\ell^+\ell^-$ decay compared with those of $B\to
K^*\ell^+\ell^-$ decay are more sensitive to the dileptonic
invariant mass.
\end{abstract}
\end{titlepage}
\section{Introduction}
The rare flavor-changing neutral-current(FCNC) processes are used to
test the predictions of Standard Model(SM) at loop level and for
searching new-physics(NP). In this regard, $b \to s(d)$ and $\mu \to
e$ transitions have been  studied to check predictions of SM at loop
level and to look at the NP via their indirect effects where the
direct productions are not accessible at present  collider
experiments. Semileptonic and radiative $B$ decays involving a
vector or axial vector meson have been observed by BABAR, BELLE and
CLEO. For $B\to K^\star \ell^+ \ell^-$ decays, the forward-backward
asymmetry has been measured by BABAR~\cite{Aubert:2006vb} and
BELLE~\cite{Ishikawa:2006fh}. Recently, BABAR
\cite{aubert:2008ju,Aubert:2008ps,Eigen:2008nz} has reported the
measurements for the longitudinal polarization fraction and
forward-backward asymmetry (FBA) of $B\to K^*(892) \ell^+ \ell^-$,
and for the isospin asymmetry of $B^0\to K^{*0}(892) \ell^+ \ell^-$
and $B^\pm\to K^{*\pm}(892) \ell^+ \ell^-$ channels. The data may
challenge the sings of Wilson coefficients, for instance,
$C^{eff}_7$. In order to extract  the magnitudes and  arguments of
the effective Wilson coefficients, one may  measure various
observables in various inclusive and exclusive rare processes. In
this regrad, the studies of asymmetries, which are less sensitive to
the hadronic uncertainties than the branching ratio, are favored.
The studies of inclusive and exclusive rare processes as well as
various asymmetries should be considerably improved at LHCb. The
radiative $B$ decay involving the $\konel$, the orbitally excited
($P$-wave) state, is recently observed by \BELLE\ and other
radiative and semileptonic decay modes involving $\konel$ and
$\koneh$ are hopefully expected to be observed soon. Some studies
for $B\to\kone\lpm$ involving formfactors, branching ratio and
forward-backward(FB) asymmetries of semileptonic decay modes have
been made recently
\cite{Paracha:2007yx,Ahmed:2008ti,Saddique:2008xj,Hatanaka:2008gu}.
In present work, we study the single and double lepton polarization
asymmetries in the $B \to K_1(1270) \ell^+ \ell^-$ decays. These
studies are  complimentary to the studies of branching ratio and FB
asymmetries. Note that, just like $B \to \Kstar\lpm$ decays
\cite{Ali:1999mm,Ali:2002jg,Beneke:2001at,Feldmann:2002iw,Kruger:2005ep,Bobeth:2008ij,Egede:2008uy,Altmannshofer:2008dz,Chen:2008ug,VBKZPRD},
$B\to\kone\lpm$ decays can be studied for the NP effects, however,
these are much more sophisticated due to the mixing of the $\konea$
and $\koneb$, which are the $\sanpone$ and $\onepone$ states,
respectively.  The physical $K_1$ mesons are $\konel$ and $\koneh$,
described by\cite{Hatanaka:2008gu}
\begin{eqnarray}
\pmatrix{|\barkonel \rangle \cr |\barkoneh \rangle} = M \pmatrix{|
\barkonea \rangle \cr | \barkoneb \rangle}, \quad\mbox{with} \quad M
= \pmatrix{ \sin \thetaK & \phantom{-} \cos \thetaK \cr \cos \thetaK
& -\sin \thetaK}. \label{mixing}
\end{eqnarray}

The mixing angle  $\thetaK$ was estimated to be $|\thetaK| \approx
34\degree\vee 57\degree$ in Ref. \cite{Suzuki:1993yc}, $35\degree
\leq |\thetaK| \leq 55\degree$ in Ref.~\cite{Burakovsky:1997ci},
$|\thetaK|= 37\degree \vee 58\degree$ in Ref.~\cite{Cheng:2003bn},
and $\thetaK= -(34 \pm 13)\degree $ in \cite{Hatanaka:2008gu,
Hatanaka:2008xj}. In this study we will use the results of
Ref.\cite{Hatanaka:2008gu,Hatanaka:2008xj} for numerical
calculations.

The paper includes 5 sections: In section 2, we recall  the
effective Hamiltonian for $B \to K_1(1270) \ell^+ \ell^-$ decays. In
section 3 we recall the calculations of effective Hamiltonian.  In
section 4, single and double lepton polarization asymmetries are
derived, respectively. In section 5, we examine the sensitivity of
these physical observable to the invariant dileptonic mass and our
conclusion.

\section{The effective Hamiltonian}\label{sec:Hamiltonian}
Using the QCD corrected effective Hamiltonian, the matrix element
$b\rightarrow s\ell^{+}\ell^{-}$ can be written as:

\begin{eqnarray}
M(b \rightarrow s\ell ^{+}\ell ^{-})=\frac{G_{F}\alpha }{\sqrt{2}%
\pi }V_{tb}V_{td}^{\ast }\left\{
\begin{array}{c}
c_{9}^{eff}\left[ \bar{d}\gamma _{\mu }Lb\right] \left[
\bar{\ell}\gamma
^{\mu }\ell \right] \\
+c_{10}\left[ \bar{d}\gamma _{\mu }Lb\right] \left[ \bar{\ell}\gamma
^{\mu
}\gamma ^{5}\ell \right] \\
-2\hat{m}_{b}c_{7}^{eff}\left[ \bar{d}i\sigma _{\mu \nu }\frac{\hat{q}^{\nu }%
}{\hat{s}}Rb\right] \left[ \bar{\ell}\gamma ^{\mu }\ell \right]%
\end{array}
\right\}  \nonumber \\
&&  \label{e1}
\end{eqnarray}
where $c_{i}$ are Wilson coefficients calculated in naive
dimensional regularization~(NDR) scheme at the leading order(LO),
next to leading order(NLO) and next-to-next leading order (NNLO) in
the SM\cite{Buras:1994dj}--\cite{NNLL}. $c_9^\eff(\hats) = c_9 +
Y(\hats)$, where $Y(\hats) = Y_{\rm pert}(\hats) + Y_{\rm LD}$
contains both the perturbative part $Y_{\rm pert}(\hats)$ and
long-distance part $Y_{\rm LD}(\hats)$.
$Y(\hats)_{\rm pert}$ is given by \cite{Buras:1994dj}
\begin{eqnarray}
Y_{\rm pert} (\hats) &=& g(\hatm_c,\hats) c_0 \nonumber\\&&
-\frac{1}{2} g(1,\hats) (4 \barc_3 + 4 \barc_4 + 3 \barc_5 +
\barc_6) -\frac{1}{2} g(0,\hats) (\barc_3 + 3 \barc_4) \nonumber\\&&
+\frac{2}{9} (3 \barc_3 + \barc_4 + 3 \barc_5 + \barc_6),
\\
\mbox{with}\quad c_0 &\equiv& \barc_1 + 3\barc_2 + 3 \barc_3 +
\barc_4 + 3 \barc_5 + \barc_6,
\end{eqnarray}
and the function $g(x,y)$ defined in \cite{Buras:1994dj}.  Here,
$\barc_1$ -- $\barc_6$ are the Wilson coefficients in the leading
logarithmic approximation. The relevant Wilson coefficients are
given  in Refs.~\cite{Ali:1999mm}. $Y(\hats)_{\rm LD}$ involves $B
\to K_1 V(\cbar c)$ resonances \cite{Lim:1988yu,
Ali:1991is,Kruger:1996cv}, where $V(\cbar c)$ are the vector
charmonium states. We follow Refs.~\cite{Lim:1988yu,Ali:1991is} and
set
\begin{eqnarray}
Y_{\rm LD}(\hats) &=&
 - \frac{3\pi}{\alphaem^2} c_0
\sum_{V = \psi(1s),\cdots} \kappa_V \frac{\hatm_V \Br(V\to
\l^+\l^-)\hat{\Gamma}_{\rm tot}^V}{\hats - \hatm_V^2 + i \hatm_V
\hat{\Gamma}_{\rm tot}^V},
\end{eqnarray}
where $\hat{\Gamma}_{\rm tot}^V \equiv \Gamma_{\rm tot}^V/\mB$ and
$\kappa_V$ takes different value for different exclusive
semileptonic decay. This phenomenological parameters $\kappa_V$ can
be fixed for $B\rightarrow K^\ast \ell^+\ell^-$ decay by equating
the naive factorization estimate of the $B\rightarrow K^\ast V$ rate
and the experimental measured results\cite{Ali:1999mm}. Except for
the branching ratio of $B\rightarrow J/\Psi
K_1(1270)$\cite{Yao:2006px}, there is no experimental results on $B
\rightarrow K_1V (c\bar{c})$. Thus, we will use the results of
$B\rightarrow K^\ast V$ to estimate the values of $\kappa_V$. We
assume that the effect of substituting $K^\ast$ with $K_1$ is
identical in the radiative and in the non leptonic decay, in other
words that each form factor for the $B\rightarrow K_1$ transition is
given by the corresponding form factor for $B\rightarrow K^\ast$
multiplied by the same factor y, which is define as
follows\cite{Nardulli:2005fn}: \beq y=\frac{f^{B\rightarrow
K_1}(0)}{f^{B\rightarrow K^\ast}(0)}\approx 1.06 \eeq once the
change of parity between the two strange mesons is taken into
account. We predict that \bea\label{ratio} \kappa_V(B\rightarrow
K_1)\approx 1.06 ~\kappa_V(B\rightarrow K^\ast)\eea. Using the above
equation and the results for $\kappa_V$ obtained for $B\rightarrow
K^\ast$ transition\cite{Ali:1999mm}. We find $\kappa_V=1.75$ for
$J/\Psi(1S)$ and $\kappa_V=2.43$ for $\Psi(2S)$, respectively. The
relevant properties of vector charmonium states are summarized in
Table~\ref{charmonium}.
\begin{table}[tbp]
\caption{Masses, total decay widths and branching fractions of
dilepton decays of vector charmonium states
\cite{Yao:2006px}.}\label{charmonium}
\begin{center}
\begin{tabular}{cclll}
$V$ & Mass[\GeV] &  $\Gamma_{\rm tot}^V$[\MeV]
 &\multicolumn{2}{c}{$\Br(V\to\lpm)$}
\\
\hline $J/\Psi(1S)$ & $3.097$ & $0.093$ & $5.9\times10^{-2}$ & for
$\l=e,\mu$
\\
$\Psi(2S)$   & $3.686$ & $0.327$ & $7.4\times10^{-3}$ & for $\l=e,\mu$ \\
             &         &             & $3.0\times10^{-3}$ & for $\l=\tau$
\\
$\Psi(3770)$ & $3.772$ & $25.2$ & $9.8\times10^{-6}$ & for $\l=e$
\\
$\Psi(4040)$ & $4.040$ & $80$ & $1.1\times10^{-5}$ & for $\l=e$
\\
$\Psi(4160)$ & $4.153$ & $103$ & $8.1\times10^{-6}$ & for $\l=e$
\\
$\Psi(4415)$ & $4.421$ & $62$ & $9.4\times10^{-6}$ & for $\l=e$
\end{tabular}
\end{center}
\end{table}

The matrix element for the exclusive decay can be obtain by
sandwiching Eq.~(\ref{e1}) between initial hadron state $B(p_B)$ and
final hadron state $K_1$ in terms of formfactors.

The $\barB(p_B)\to \barkone(\pA,\lambda)$ formfactors are defined as
(see \cite{Hatanaka:2008gu})
\begin{eqnarray}
\lefteqn{\bra{\barkone(\pA,\lambda)}|\bar{s} \gamma_\mu (1-\gamma_5)
b|\ket{\barB(p_B)}}
\quad&&\nonumber\\
&=&
 -i \frac{2}{m_B + \mkone} \epsilon_{\mu\nu\rho\sigma}
\varepsilon_{(\lambda)}^{*\nu} p_B^\rho \pA^\sigma A^{\kone}(q^2)
\nonumber
\\
&& -\left[ (m_B + \mkone)\varepsilon_\mu^{(\lambda)*}
V_1^{\kone}(q^2) - (p_B + \pA)_\mu (\varepsilon_{(\lambda)}^* \cdot
p_B)
 \frac{V_2^{\kone}(q^2)}{m_B + \mkone}
\right] \nonumber\\&& +2 \mkone \frac{\varepsilon_{(\lambda)}^*
\cdot p_B}{q^2} q_\mu \left[
 V_3^{\kone}(q^2) - V_0^{\kone}(q^2)
\right], \label{formfactor1}
\\
\lefteqn{\bra{\barkone(\pA,\lambda)}|\bar{s} \sigma_{\mu\nu} q^\nu
(1+\gamma_5) b|\ket{\barB(p_B)}}
\quad&&\nonumber\\
&=& 2T_1^{\kone}(q^2) \epsilon_{\mu\nu\rho\sigma}
\varepsilon_{(\lambda)}^{*\nu} p_B^\rho  \pA^\sigma
\nonumber\\
&& -i T_2^{\kone}(q^2) \left[
 (m_B^2 - \mkone^2) \varepsilon^{(\lambda)}_{*\mu}
-(\varepsilon_{(\lambda)}^{*}\cdot q)
 (p_B + \pA)_\mu
\right] \nonumber\\&& - iT_3^{\kone}(q^2)
(\varepsilon_{(\lambda)}^{*} \cdot q) \left[ q_\mu -
\frac{q^2}{m_B^2 - \mkone^2} (\pA + p_B)_\mu \right],
\label{formfactor2}
\end{eqnarray}
where $q \equiv p_B - \pA=p_{\ell^+}+p_{\ell^-}$.  In order to
ensure finiteness  at $q^2=0$, it is  required
\begin{eqnarray}
V_3^\kone(0) &=& V_0^{\kone}(0), \quad T_1^{\kone}(0) =
T_2^{\kone}(0),
\nonumber \\
V_3^\kone(q^2) &=& \frac{m_B + m_{\kone}}{2 m_{\kone}}
V_1^\kone(q^2)-
           \frac{m_B - m_{\kone}}{2 m_{\kone}} V_2^\kone(q^2).
\end{eqnarray}
The formfactors of $B\rightarrow K_1(1270)$ and $B\rightarrow
K_1(1400)$ can be expressed in terms of $B\rightarrow K_A$ and
$B\rightarrow K_B$ as follows(see \cite{Hatanaka:2008gu}):
\begin{eqnarray}
\pmatrix{
 \bra{\barkonel}|\sbar \gamma_\mu(1-\gamma_5) b |\ket{\barB} \cr
 \bra{\barkoneh}|\sbar \gamma_\mu(1-\gamma_5) b |\ket{\barB}}
&=& M \pmatrix{
 \bra{\barK_{1A}}|\sbar\gamma_\mu(1-\gamma_5) b|\ket{\barB} \cr
 \bra{\barK_{1B}}|\sbar\gamma_\mu(1-\gamma_5) b|\ket{\barB} },
\\
\pmatrix{ \bra{\barkonel}|\sbar \sigma_{\mu\nu}q^\nu(1+\gamma_5) b
|\ket{\barB} \cr \bra{\barkoneh}|\sbar
\sigma_{\mu\nu}q^\nu(1+\gamma_5) b |\ket{\barB} } &=& M \pmatrix{
 \bra{\barK_{1A}}|\sbar\sigma_{\mu\nu}q^\nu(1+\gamma_5) b|\ket{\barB} \cr
 \bra{\barK_{1B}}|\sbar\gamma_{\mu\nu}q^\nu(1+\gamma_5) b|\ket{\barB} }
,
\end{eqnarray}
using  the mixing matrix $M$ being given in Eq.~\eqref{mixing} the
formfactors $A^\kone,V_{0,1,2}^\kone$ and $T_{1,2,3}^\kone$ can be
written as follows:
\begin{eqnarray}
\pmatrix{
 A^{\konel}/(m_B + m_{\konel}) \cr
 A^{\koneh}/(m_B + m_{\koneh})}
&=& M \pmatrix{
 A^{\konea}/(m_B + m_{\konea}) \cr
 A^{\koneb}/(m_B + m_{\koneb})},
\\
\pmatrix{ (m_B+m_{\konel}) V_1^{K_1(1270)} \cr (m_B+m_{\koneh})
V_1^{K_1(1400)}} &=& M \pmatrix{ (m_B+m_{\konea})V_1^{K_{1A}} \cr
(m_B+m_{\koneb})V_1^{K_{1B}}},
\\
\pmatrix{ V_2^{K_1(1270)}/(m_B + m_{\konel}) \cr
V_2^{K_1(1400)}/(m_B + m_{\koneh})} &=& M \pmatrix{
V_2^{K_{1A}}/(m_B + m_{\konea}) \cr V_2^{K_{1B}}/(m_B +
m_{\koneb})},
\\
\pmatrix{ m_{\konel} V_0^{K_1(1270)} \cr m_{\koneh} V_0^{K_1(1400)}}
&=& M \pmatrix{ m_{\konea} V_0^{K_{1A}} \cr m_{\koneb}
V_0^{K_{1B}}},
\\
\pmatrix{ T_1^{K_1(1270)} \cr T_1^{K_1(1400)}}&=& M \pmatrix{
T_1^{K_{1A}} \cr T_1^{K_{1B}} },
\\
\pmatrix{ (m_B^2 - m_{\konel}^2) T_2^{K_1(1270)} \cr (m_B^2 -
m_{\koneh}^2) T_2^{K_1(1400)}}&=& M \pmatrix{ (m_B^2 - m_{\konea}^2)
T_2^{K_{1A}} \cr (m_B^2 - m_{\koneb}^2) T_2^{K_{1B}} },
\\
\pmatrix{ T_3^{K_1(1270)} \cr T_3^{K_1(1400)}} &=& M \pmatrix{
T_3^{K_{1A}} \cr T_3^{K_{1B}} },
\end{eqnarray}
where it is supposed that $p^\mu_{\konel,\koneh} \simeq
p^\mu_{\konea} \simeq p^\mu_{\koneb}$\cite{Hatanaka:2008gu}. These
formfactors within light-cone sum rule (LCSR) are estimated in
\cite{Yang:2008xw}. The momentum dependence of all formfactors is
parameterized as:
\begin{eqnarray} \label{eq:FFpara}
 F(q^2)=\,{F(0)\over 1-a(q^2/m_{B}^2)+b(q^2/m_{B}^2)^2}.
\end{eqnarray}

The values of $F(0)$, $a$  and $b$ parameters are exhibited in
Table~\ref{tab:FFinLF}.

Thus, the matrix element for $B\to\kone\lpm$ in terms of formfactos
is given by
\begin{eqnarray}
{\cal M} &=& \frac{G_F \alpha_{\rm em}}{2\sqrt{2}\pi} V_{ts}^*
V_{tb}^{}\, m_B \cdot (-i)\left[
  \T_\mu^{(\kone),1} \lbar \gamma^\mu \l
 +\T_\mu^{(\kone),2} \lbar \gamma^\mu \gamma_5 \l
\right],
\end{eqnarray}

where
\begin{eqnarray}
\T_\mu^{(\kone),1} &=&
 \A^\kone(\hats) \epsilon_{\mu\nu\rho\sigma}
 \varepsilon^{*\nu} \hatp_B^\rho \hatp_\kone^\sigma
-i \B^\kone(\hats)\varepsilon^{*}_\mu \nonumber\\&& +i
\C^\kone(\hats)( \varepsilon^{*} \cdot \hatp_B) \hatp_\mu +i
\D^\kone(\hats)( \varepsilon^{*} \cdot \hatp_B) \hatq_\mu,
\\
\T_\mu^{(\kone),2} &=&
 \E^\kone(\hats) \epsilon_{\mu\nu\rho\sigma}
 \varepsilon^{*\nu} \hatp_B^\rho \hatp_\kone^\sigma
-i \F^\kone(\hats)\varepsilon^{*}_\mu \nonumber\\&& +i
\G^\kone(\hats)( \varepsilon^{*} \cdot \hatp_B) \hatp_\mu +i
\H^\kone(\hats)( \varepsilon^{*} \cdot \hatp_B) \hatq_\mu,
\end{eqnarray}
with
 $\hatp = p/m_B$,
 $\hatp_B = p_B/m_B$,
 $\hatq= q/m_B$ and
 $p = p_B + p_\kone$,
 $q = p_B - p_\kone = p_{\ell^+} + p_{\ell^-} $.

Here $\A^\kone(\hats), \cdots, \H^\kone(\hats)$ are defined by
\begin{eqnarray}
\A^\kone(\hats) &=& \frac{2}{1+\sqrt{\hat{r}_{K_1}}} c_9^{\eff}
(\hats) A^\kone(\hats) + \frac{4\hatm_b}{\hats} c_7^\eff
T^\kone_1(\hats), \label{Eq:A}
\\
\B^\kone(\hats) &=& (1+\sqrt{\hat{r}_{K_1}})\left[
 c_9^\eff (\hats) V_1^\kone(\hats)
 + \frac{2\hatm_b}{\hats} (1-\sqrt{\hat{r}_{K_1}})c_7^\eff T^\kone_2(\hats)
\right],
\\
\C^\kone(\hats) &=& \frac{1}{1-\hat{r}_{K_1}} \left[
 (1-\sqrt{\hat{r}_{K_1}}) c_9^\eff(\hats) V_2^\kone (\hats) + 2\hatm_b c_7^\eff
 \left(
  T_3^\kone(\hats) + \frac{1-\sqrt{\hat{r}_{K_1}}^2}{\hats} T_2^\kone(\hats)
 \right)
\right],
\nonumber\\
\\
\D^\kone(\hats) &=& \frac{1}{\hats} \biggl[
 c_9^\eff(\hats) \left\{(1+\sqrt{\hat{r}_{K_1}}) V_1^\kone(\hats)
  - (1-\sqrt{\hat{r}_{K_1}}) V_2^\kone(\hats)
  - 2\sqrt{\hat{r}_{K_1}} V_0^\kone(\hats) \right\}
\nonumber\\&&
  - 2\hatm_b c_7^\eff T_3^\kone(\hats)
\biggr],
\\
\E^\kone(\hats) &=& \frac{2}{1+\sqrt{\hat{r}_{K_1}}} c_{10}
A^\kone(\hats),
\\
\F^\kone(\hats) &=& (1 + \sqrt{\hat{r}_{K_1}}) c_{10}
V_1^\kone(\hats),
\\
\G^\kone(\hats) &=& \frac{1}{1 + \sqrt{\hat{r}_{K_1}}} c_{10}
V_2^\kone(\hats),
\\
\H^\kone(\hats) &=& \frac{1}{\hats} c_{10} \left[
 (1+\sqrt{\hat{r}_{K_1}}) V_1^\kone(\hats)
 - (1-\sqrt{\hat{r}_{K_1}}) V_2^\kone(\hats) - 2\sqrt{\hat{r}_{K_1}} V_0^\kone(\hats)
\right], \label{Eq:H}
\end{eqnarray}
with $\hat{r}_{K_1} = m^2_\kone/m^2_B$, $\hat{m_{\ell}}=m_\ell/m_B$
and $\hats = q^2/m_B^2$.
\begin{table}[t]
\caption{Formfactors for $B\to K_{1A},K_{1B}$ transitions obtained
in the LCSR calculation \cite{Yang:2008xw} are fitted to the
3-parameter form in Eq. (\ref{eq:FFpara}).} \label{tab:FFinLF}
\begin{tabular}{clll|clll}
      ~~~~$F$~~~~~~
    & ~~~~~$F(0)$~~~~~
    & ~~~$a$~~~
    & ~~~$b$~~
    & ~~~~$F$~~~~~~
    & ~~~~~$F(0)$~~~~~
    & ~~~$a$~~~
    & ~~~$b$~~
 \\
    \hline
$V_1^{BK_{1A}}$
    & $0.34\pm0.07$
    & $0.635$
    & $0.211$
&$V_1^{BK_{1B}}$
    & $-0.29^{+0.08}_{-0.05}$
    & $0.729$
    & $0.074$
    \\
$V_2^{BK_{1A}}$
    & $0.41\pm 0.08$
    & $1.51$
    & $1.18~~$
&$V_2^{BK_{1B}}$
    & $-0.17^{+0.05}_{-0.03}$
    & $0.919$
    & $0.855$
    \\
$V_0^{BK_{1A}}$
    & $0.22\pm0.04$
    & $2.40$
    & $1.78~~$
&$V_0^{BK_{1B}}$
    & $-0.45^{+0.12}_{-0.08}$
    & $1.34$
    & $0.690$
    \\
$A^{BK_{1A}}$
    & $0.45\pm0.09$
    & $1.60$
    & $0.974$
&$A^{BK_{1B}}$
    & $-0.37^{+0.10}_{-0.06}$
    & $1.72$
    & $0.912$
    \\
$T_1^{BK_{1A}}$
    & $0.31^{+0.09}_{-0.05}$
    & $2.01$
    & $1.50$
&$T_1^{BK_{1B}}$
    & $-0.25^{+0.06}_{-0.07}$
    & $1.59$
    & $0.790$
    \\
$T_2^{BK_{1A}}$
    & $0.31^{+0.09}_{-0.05}$
    & $0.629$
    & $0.387$
&$T_2^{BK_{1B}}$
    & $-0.25^{+0.06}_{-0.07}$
    & $0.378$
    & $-0.755$
    \\
$T_3^{BK_{1A}}$
    & $0.28^{+0.08}_{-0.05}$
    & $1.36$
    & $0.720$
&$T_3^{BK_{1B}}$
    & $-0.11\pm 0.02$
    & $-1.61$
    & $10.2$
\end{tabular}

\end{table}
The differential decay spectrum can be obtained from the decay
amplitude
\begin{eqnarray}
\frac{d \Gamma(\barB\to\barkone\lpm)}{d \hats} = \frac{G_F^2
\alphaem^2 m_B^5}{2^{8}\pi^5}
 \left|V_{tb}V_{ts}^*\right|^2 v\sqrt{\lambda} \Delta(\hat{s})
\end{eqnarray}

\bea\label{dgds1}\nnb \Delta&=&\frac{8Re[{\cal F}{\cal H}^*]
\hat{m}_{\ell}^2 \lambda}{\hat{r}_{K_1}}+\frac{8Re[{\cal G}{\cal
H}^*]
\hat{m}_{\ell}^2(-1+\hat{r}_{K_1})\lambda}{\hat{r}_{K_1}}-\frac{8|{\cal
H}|^2 \hat{m}_{\ell}^2 \hat{s} \lambda}{\hat{r}_{K_1}}
\\ \nnb &-&
\frac{2Re[{\cal B}{\cal
C}^*](-1+\hat{r}_{K_1}+\hat{s})(3+3\hat{r}_{K_1}^2-6\hat{s}+3\hat{s}^2-6\hat{r}_{K_1}(1+\hat{s})-v^2\lambda)}{3\hat{r}_{K_1}}
\\ \nnb &-&
\frac{|{\cal C}|^2
\lambda(3+3\hat{r}_{K_1}^2-6\hat{s}+3\hat{s}^2-6\hat{r}_{K_1}(1+\hat{s})-v^2\lambda)}{3\hat{r}_{K_1}}
\\ \nnb &-&
\frac{|{\cal G}|^2
\lambda(3+3\hat{r}_{K_1}^2+12\hat{m}_{\ell}^2(2+2\hat{r}_{K_1}-\hat{s})-6\hat{s}+3\hat{s}^2-6\hat{r}_{K_1}(1+\hat{s})-v^2\lambda)}{3\hat{r}_{K_1}}
\\ \nnb &+&
\frac{|{\cal
F}|^2(-3-3\hat{r}_{K_1}^2+6\hat{r}_{K_1}(1+16\hat{m}_{\ell}^2-3\hat{s})+6\hat{s}-3\hat{s}^2+v^2\lambda)}{3\hat{r}_{K_1}}
\\ \nnb &+&
\frac{|{\cal
B}|^2(-3-3\hat{r}_{K_1}^2+6\hat{s}-3\hat{s}^2-6\hat{r}_{K_1}(-1+8\hat{m}_{\ell}^2+3\hat{s})+v^2\lambda)}{3\hat{r}_{K_1}}
\\ \nnb &+&
\frac{2}{3\hat{r}_{K_1}}Re[{\cal F}{\cal
G}^*](12\hat{m}_{\ell}^2\lambda-(-1+\hat{r}_{K_1}+\hat{s})
(3+3\hat{r}_{K_1}^2-6\hat{s}+3\hat{s}^2-6\hat{r}_{K_1}(1+\hat{s})-v^2\lambda))
\\ \nnb &+&
|{\cal
A}|^2(-4\hat{m}_{\ell}^2\lambda-\frac{\hat{s}}{3}(3+3\hat{r}_{K_1}^2-6\hat{s}+3\hat{s}^2-6\hat{r}_{K_1}(1+\hat{s})+v^2\lambda))
\\ \nnb &+&
|{\cal
E}|^2(4\hat{m}_{\ell}^2\lambda-\frac{\hat{s}}{3}(3+3\hat{r}_{K_1}^2-6\hat{s}+3\hat{s}^2-6\hat{r}_{K_1}(1+\hat{s})+v^2\lambda))
 \eea
\section{Lepton polarization asymmetries}

In order to calculate the polarization asymmetries of  the leptons,
 we must first define the orthogonal vectors $S$ in
the rest frame of $\ell^-$ and $W$ in the rest frame of $\ell^+$
(where these vectors are the polarization vectors of the leptons).
Note that, we will use the subscripts $L$, $N$ and $T$ to correspond
to the leptons which are polarized along with the longitudinal,
normal and transverse polarization of leptons, respectively.
\cite{Kruger:1996cv, Fukae}.
\begin{eqnarray}
S^\mu_L & \equiv & (0, \mathbf{e}_{L}) ~=~ \left(0,
\frac{\mathbf{p}_-}{|\mathbf{p}_-|}
\right) , \nonumber \\
S^\mu_N & \equiv & (0, \mathbf{e}_{N}) ~=~ \left(0,
\frac{\mathbf{p}_{K_1} \times \mathbf{p}_-}{|\mathbf{p}_{K_1} \times
\mathbf{p}_- |}\right) , \nonumber \\
S^\mu_T & \equiv & (0, \mathbf{e}_{T}) ~=~ \left(0, \mathbf{e}_{N}
\times \mathbf{e}_{L}\right) , \label{sec3:eq:1} \\
W^\mu_L & \equiv & (0, \mathbf{w}_{L}) ~=~ \left(0,
\frac{\mathbf{p}_+}{|\mathbf{p}_+|} \right) , \nonumber \\
W^\mu_N & \equiv & (0, \mathbf{w}_{N}) ~=~ \left(0,
\frac{\mathbf{p}_{K_1} \times \mathbf{p}_+}{|\mathbf{p}_{K_1} \times
\mathbf{p}_+ |} \right) , \nonumber \\
W^\mu_T & \equiv & (0, \mathbf{w}_{T}) ~=~ (0, \mathbf{w}_{N} \times
\mathbf{w}_{L}) , \label{sec3:eq:2}
\end{eqnarray}
where $\mathbf{p}_+$, $\mathbf{p}_-$ and $\mathbf{p}_{K_1}$ are the
three momenta of the $\ell^+$, $\ell^-$ and $K_1$ particles,
respectively. On boosting the vectors defined by
Eqs.~(\ref{sec3:eq:1},\ref{sec3:eq:2}) to the CM frame of the
$\ell^- \ell^+$ system only the longitudinal vector will be boosted,
while the other two remain the same. The longitudinal vectors in the
CM frame of the $\ell^- \ell^+$ system  become;
\begin{eqnarray}
S^\mu_L & = & \left( \frac{|\mathbf{p}_-|}{m_\ell}, \frac{E_{\ell}
\mathbf{p}_-}{m_\ell |\mathbf{p}_-|} \right) , \nonumber \\
W^\mu_L & = & \left( \frac{|\mathbf{p}_-|}{m_\ell}, - \frac{E_{\ell}
\mathbf{p}_-}{m_\ell |\mathbf{p}_-|} \right) . \label{sec3:eq:3}
\end{eqnarray}
The polarization asymmetries can now be calculated using the spin
projector ${1 \over 2}(1 + \gamma_5 \!\!\not\!\! S)$ for $\ell^-$
and the spin projector ${1 \over 2}(1 + \gamma_5\! \not\!\! W)$ for
$\ell^+$. The single and double--lepton polarization asymmetries $
P_{ij}$ are defined, respectively, as \cite{Fukae}
\beq  P_{i} =\frac{\frac{d\Gamma({\bf s^{\pm}}={\bf
\hat{i}})}{d\hat{s}}-\frac{d\Gamma({\bf s^{\pm}}={-\bf
      \hat{i}})}{d\hat{s}}}{\frac{d\Gamma({\bf s^{\pm}}={\bf \hat{i}})}{d\hat{s}}+\frac{d\Gamma({\bf s^{\pm}}={-\bf
      \hat{i}})}{d\hat{s}}}
\eeq
 and
 \beq  P_{ij} =\frac{\Big[\frac{d\Gamma({\bf s^{+}}={\bf
\hat{i}},{\bf
      s^{-}}={\bf \hat{j}})}{d\hat{s}}-\frac{d\Gamma({\bf s^{+}}={\bf
      \hat{i}},{\bf s^{-}}={-\bf \hat{j}})}{d\hat{s}}\Big]
  -\Big[\frac{d\Gamma({\bf s^{+}}={-\bf \hat{i}},{\bf s^{-}}={\bf
      \hat{j}})}{d\hat{s}}-\frac{d\Gamma({\bf s^{+}}={-\bf
      \hat{i}},{\bf s^{-}}={-\bf \hat{j}})}{d\hat{s}}\Big]}
{\Big[\frac{d\Gamma({\bf s^{+}}={\bf \hat{i}},{\bf s^{-}}={\bf
      \hat{j}})}{d\hat{s}}+\frac{d\Gamma({\bf s^{+}}={\bf
      \hat{i}},{\bf s^{-}}={-\bf \hat{j}})}{d\hat{s}}\Big]
  +\Big[\frac{d\Gamma({\bf s^{+}}={-\bf \hat{i}},{\bf s^{-}}={\bf
      \hat{j}})}{d\hat{s}}+\frac{d\Gamma({\bf s^{+}}={-\bf
      \hat{i}},{\bf s^{-}}={-\bf \hat{j}})}{d\hat{s}}\Big]},
\eeq
where ${\hat i}=L,N,T$ and ${\hat j}=L,N,T$ are unit vectors.

Equipped with these definitions, we evaluate the single and double
lepton polarization asymmetries and obtain the following results:

\bea\label{polL} P_L&=&-\frac{2 Re[{\cal B G^\ast}]
(\hat{r}_{K_1}+\hat{s}-1) v \left(3 \hat{r}_{K_1}^2-6 (\hat{s}+1)
\hat{r}_{K_1}+3 (\hat{s}-1)^2-\lambda \right)}{3 \hat{r}_{K_1}}\nnb
\\&-&\frac{2 Re[{\cal CF^\ast}] (\hat{r}_{K_1}+\hat{s}-1) v \left(3 \hat{r}_{K_1}^2-6 (\hat{s}+1) \hat{r}_{K_1}+3 (\hat{s}-1)^2-\lambda \right)}{3 \hat{r}_{K_1}}\nnb \\
&-&\frac{2 Re[{\cal CG^\ast}] v \lambda  \left(3 \hat{r}_{K_1}^2-6
(\hat{s}+1) \hat{r}_{K_1}+3 (\hat{s}-1)^2-\lambda \right)}{3
\hat{r}_{K_1}}\nnb\\&-&\frac{2}{3}
   Re[{\cal AE^\ast}] \hat{s} v \left(3 \hat{r}_{K_1}^2-6 (\hat{s}+1) \hat{r}_{K_1}+3 (\hat{s}-1)^2+\lambda \right)\nnb \\&+&\frac{2 Re[{\cal BF^\ast}] v
   \left(\lambda -3 \left(\hat{r}_{K_1}^2+(6 \hat{s}-2) \hat{r}_{K_1}+(\hat{s}-1)^2\right)\right)}{3 \hat{r}_{K_1}}
\eea

\bea P_{T}&=&\pi  \hat{m}_{\ell}\sqrt{\lambda }\Bigg\{-\frac{
Re[{\cal CH^\ast}] \sqrt{\hat{s}} \lambda
}{\hat{r}_{K_1}}+\frac{Re[{\cal CF^\ast}]  \lambda }{\hat{r}_{K_1}
   \sqrt{\hat{s}}}+\frac{Re[{\cal CG^\ast}]  (\hat{r}_{K_1}-1)
    \lambda}{\hat{r}_{K_1} \sqrt{\hat{s}}}\nnb \\&-&\frac{Re[{\cal BH^\ast}]  \sqrt{\hat{s}}
   (\hat{r}_{K_1}+\hat{s}-1) }{\hat{r}_{K_1}}+\frac{Re[{\cal BF^\ast}]
   (\hat{r}_{K_1}+\hat{s}-1) }{\hat{r}_{K_1}
   \sqrt{\hat{s}}}\nnb \\&+&\frac{Re[{\cal BG^\ast}]  (\hat{r}_{K_1}-1) (\hat{r}_{K_1}+\hat{s}-1)
    }{\hat{r}_{K_1} \sqrt{\hat{s}}}-4 Re[{\cal AB^\ast}]
   \sqrt{\hat{s}} \Bigg\}\nnb\\
P_N&=& i \pi \hat{m}_{\ell}\sqrt{\lambda }\Bigg\{\frac{ Im[{\cal
GH^\ast}]  \sqrt{\hat{s}}  \lambda}{\hat{r}_{K_1}}+\frac{ Im[{\cal
FG^\ast}]
 \sqrt{\hat{s}} (-3
   \hat{r}_{K_1}+\hat{s}-1) }{\hat{r}_{K_1}}
   \\&+&\frac{ Im[{\cal FH^\ast}]  \sqrt{\hat{s}}
    (\hat{r}_{K_1}+\hat{s}-1)}{\hat{r}_{K_1}}-2  Im[{\cal AF^\ast}]  \sqrt{\hat{s}}
   -2  Im[{\cal AE^\ast}]  \sqrt{\hat{s}}\Bigg\}\nnb
\eea

\bea\label{polLL2}\nnb P_{LL}&=&\frac{4Re[{\cal F}{\cal
H}^*](2\hat{m}_{\ell}^2+\hat{s})\lambda}{\hat{r}_{K_1}}+\frac{4Re[{\cal
G}{\cal
H}^*](-1+\hat{r}_{K_1})(2\hat{m}_{\ell}^2+\hat{s})\lambda}{\hat{r}_{K_1}}
-\frac{2|{\cal H}|^2 \hat{s}(2\hat{m}_{\ell}^2+\hat{s})\lambda}{\hat{r}_{K_1}}\\
\nnb &-& \frac{|{\cal G}|^2}{6\hat{m}_{\ell}^2\hat{r}_{K_1}
\hat{s}}\Bigg(\hat{s}^2(3\hat{r}_{K_1}^2+3(-1+\hat{s})^2-6\hat{r}_{K_1}(1+\hat{s})-\lambda)
\\ \nnb &+&
8\hat{m}_{\ell}^4(6+6\hat{r}_{K_1}^2+3(-2+\hat{s})\hat{s}-6\hat{r}_{K_1}(2+\hat{s})-\lambda)
\\ \nnb &-&
6\hat{m}_{\ell}^2\hat{s}(1+\hat{r}_{K_1}^2+3(-2+\hat{s})\hat{s}-2(\hat{r}_{K_1}+3\hat{r}_{K_1}
\hat{s})-\lambda)\Bigg)\lambda
\\ \nnb &-&
\frac{|{\cal
E}|^2(3(8\hat{m}_{\ell}^4+2\hat{m}_{\ell}^2\hat{s}-\hat{s}^2)\lambda+(8\hat{m}_{\ell}^4-6\hat{m}_{\ell}^2\hat{s}+\hat{s}^2)\lambda)}{6\hat{m}_{\ell}^2}
\\ \nnb &+&
\frac{Re[{\cal B}{\cal
C}^*](-1+\hat{r}_{K_1}+\hat{s})(3\hat{s}(2\hat{m}_{\ell}^2)\lambda-(8\hat{m}_{\ell}^4-3\hat{m}_{\ell}^2\hat{s}+\hat{s}^2)\lambda)}{3\hat{m}_{\ell}^2\hat{r}_{K_1}
\hat{s}}
\\ \nnb &+&
\frac{|{\cal C}|^2
\lambda(3\hat{s}(2\hat{m}_{\ell}^2+\hat{s})\lambda-(8\hat{m}_{\ell}^4-2\hat{m}_{\ell}^2\hat{s}+\hat{s}^2))\lambda}{6\hat{m}_{\ell}^2\hat{r}_{K_1}
\hat{s}}
\\ \nnb &+&
\frac{|{\cal
A}|^2(-3(8\hat{m}_{\ell}^4-6\hat{m}_{\ell}^2\hat{s}+\hat{s}^2)\lambda+(8\hat{m}_{\ell}^4-2\hat{m}_{\ell}^2\hat{s}+\hat{s}^2))\lambda}{6\hat{m}_{\ell}^2}
\\ \nnb &+&
\frac{|{\cal F}|^2}{6\hat{m}_{\ell}^2\hat{r}_{K_1}
\hat{s}}\Bigg(6\hat{m}_{\ell}^2\hat{s}(\hat{r}_{K_1}^2+(-1+\hat{s})^2+\hat{r}_{K_1}(-2+6\hat{s})-\lambda)
\\ \nnb &+&
\hat{s}^2(-3\hat{r}_{K_1}^2-3(-1+\hat{s})^2+6\hat{r}_{K_1}(1+\hat{s})+\lambda)+8\hat{m}_{\ell}^4(-6(-1+\hat{r}_{K_1}+\hat{s})^2+\lambda)\Bigg)
\\ \nnb &+&
\frac{|{\cal B}|^2}{6\hat{m}_{\ell}^2\hat{r}_{K_1}
\hat{s}}\Bigg(\hat{s}^2(3\hat{r}_{K_1}^2+3(-1+\hat{s})^2-6\hat{r}_{K_1}(1+\hat{s})-\lambda)
\\ \nnb &-&
8\hat{m}_{\ell}^4(12\hat{r}_{K_1}
\hat{s}+\lambda)+2\hat{m}_{\ell}^2\hat{s}(3(\hat{r}_{K_1}^2+(-1+\hat{s})^2+2\hat{r}_{K_1}(-1+7\hat{s}))+\lambda)\Bigg)
\\ \nnb &-&
\frac{Re[{\cal F}{\cal G}^*]}{6\hat{m}_{\ell}^2\hat{r}_{K_1}
\hat{s}}\Bigg(\hat{s}^2(-1+\hat{r}_{K_1}+\hat{s})(3\hat{r}_{K_1}^2+3(-1+\hat{s})^2-6\hat{r}_{K_1}(1+\hat{s})-\lambda)
\\ \nnb &+&
8\hat{m}_{\ell}^4(6\hat{r}_{K_1}^2-9\hat{r}_{K_1}^2(2+\hat{s})+(-1+\hat{s})(6+3(-3+\hat{s})\hat{s}-\lambda)
\\ \nnb &-&
\hat{r}_{K_1}(-18+6\hat{s}+\lambda))-
6\hat{m}_{\ell}^2\hat{s}(\hat{r}_{K_1}^3+\hat{r}_{K_1}^2(-3+\hat{s})+(-1+\hat{s})(1+\hat{s}(-4+3\hat{s})-\lambda)
\\ \nnb &-&
\hat{r}_{K_1}(-3+\hat{s}(6+5\hat{s})+\lambda))\Bigg)
 \eea

 \bea\label{polTT2}\nnb
 P_{TT}&=& \frac{|{\cal E}|^2}{3}(4\hat{m}_{\ell}^2-\hat{s})(3+3\hat{r}_{K_1}^2-6\hat{s}+3\hat{s}^2-6\hat{s}(1+\hat{s})-5\lambda)
\\ \nnb &-&
\frac{8Re[{\cal F}{\cal H}^*]
\hat{m}_{\ell}^2\lambda}{\hat{r}_{K_1}}-\frac{8Re[{\cal G}{\cal
H}^*]
\hat{m}_{\ell}^2(-1+\hat{r}_{K_1})\lambda}{\hat{r}_{K_1}}+\frac{4|{\cal
H}|^2 \hat{m}_{\ell}^2\hat{s} \lambda}{\hat{r}_{K_1}}
\\ \nnb &-&
\frac{2Re[{\cal B}{\cal C}^*]
(-1+\hat{r}_{K_1}+\hat{s})}{3\hat{r}_{K_1}
\hat{s}}\Bigg(-6(1+\hat{r}_{K_1})\hat{s}^2+3\hat{s}^2+\hat{s}(3-6\hat{r}_{K_1}+3\hat{r}_{K_1}^2-5\lambda)+4\hat{m}_{\ell}^2\lambda
\Bigg)
\\ \nnb &-&
\frac{|{\cal C}|^2
\lambda(-6(1+\hat{r}_{K_1})\hat{s}^2+3\hat{s}^2+\hat{s}(3-6\hat{r}_{K_1}+3\hat{r}_{K_1}^2-5\lambda)+4\hat{m}_{\ell}^2\lambda)}{3\hat{r}_{K_1}
\hat{s}}
\\ \nnb &+&
\frac{|{\cal
F}|^2(-6(1+\hat{r}_{K_1})\hat{s}^2+3\hat{s}^2+\hat{s}(3-6\hat{r}_{K_1}+3\hat{r}_{K_1}^2-5\lambda)+20\hat{m}_{\ell}^2\lambda)}{3\hat{r}_{K_1}
\hat{s}}
\\ \nnb &+&
\frac{|{\cal G}|^2 \lambda}{3\hat{r}_{K_1}
\hat{s}}\Bigg(-6(1+2\hat{m}_{\ell}^2+\hat{r}_{K_1})\hat{s}^2+3\hat{s}^3
\\ \nnb &+&
\hat{s}(3-6\hat{r}_{K_1}+3\hat{r}_{K_1}^2+24\hat{m}_{\ell}62(1+\hat{r}_{K_1})-5\lambda)+20\hat{m}_{\ell}^2\lambda\Bigg)
\\ \nnb &+&
|{\cal
A}|^2\Bigg(\frac{\hat{s}}{3}(3+3\hat{r}_{K_1}^2-6\hat{s}+3\hat{s}^2-6\hat{r}_{K_1}(1+\hat{s})-5\lambda)
\\ \nnb &+&
\hat{m}_{\ell}^2(-4-4\hat{r}_{K_1}^2+8\hat{s}-4\hat{s}^2+8\hat{r}_{K_1}(1+\hat{s})+\frac{4\lambda}{3})\Bigg)
\\ \nnb &+&
\frac{|{\cal
B}|^2(6(1+\hat{r}_{K_1})\hat{s}^2-3\hat{s}^3-4\hat{m}_{\ell}^2\lambda+\hat{s}(-3+(6-48\hat{m}_{\ell}^2)\hat{r}_{K_1}-3\hat{r}_{K_1}^2+5\lambda))}{3\hat{r}_{K_1}
\hat{s}}
\\ \nnb &-&
\frac{2Re[{\cal F}{\cal G}^*]}{3\hat{r}_{K_1}
\hat{s}}(3(3-4\hat{m}_{\ell}^2+\hat{r}_{K_1})\hat{s}^3-3\hat{s}^4+\hat{s}(1+4\hat{m}_{\ell}^2-\hat{r}_{K_1})(3-6\hat{r}_{K_1}+3\hat{r}_{K_1}^2-5\lambda)
\\ \nnb &-&
20\hat{m}_{\ell}^2(-1+\hat{r}_{K_1})\lambda+\hat{s}^2(-9+6\hat{r}_{K_1}+3\hat{r}_{K_1}^2-24\hat{m}_{\ell}^2(1+\hat{r}_{K_1})+5\lambda))
\eea

\bea\label{polLT2}\nnb
 P_{LT}&=&-2Re[{\cal A}{\cal F}^*+{\cal B}{\cal E}^*]\pi \hat{m}_{\ell} v \sqrt{\hat{s} \lambda}-\frac{|{\cal F}|^2\pi \hat{m}_{\ell}(-1+\hat{r}_{K_1}+\hat{s})v\sqrt{\lambda}}{\hat{r}_{K_1}\sqrt{\hat{s}}}
\\ \nnb &+&
\frac{Re[{\cal F}{\cal H}^*]\pi \hat{m}_{\ell}
v(-1+\hat{r}_{K_1}+\hat{s})\sqrt{\hat{s}
\lambda}}{\hat{r}_{K_1}}+\frac{Re[{\cal F}{\cal G}^*]\pi
\hat{m}_{\ell}(-2-2\hat{r}_{K_1}^2+3\hat{s}-\hat{s}^2+\hat{r}_{K_1}(4+\hat{s}))v\sqrt{\lambda}}{\hat{r}_{K_1}\sqrt{\hat{s}}}
\\ \nnb &-&
\frac{|{\cal G}|^2 \pi
\hat{m}_{\ell}(-1+\hat{r}_{K_1})\lambda^{\frac{3}{2}}}{\hat{r}_{K_1}\sqrt{\hat{s}}}+\frac{Re[{\cal
G}{\cal H}^*]\pi \hat{m}_{\ell} v
\lambda^{\frac{3}{2}}\sqrt{\hat{s}}}{\hat{r}_{K_1}}
 \eea

\bea\label{polLT2prim}\nnb
 P_{TL}&=&2Re[{\cal A}{\cal F}^*+{\cal B}{\cal E}^*]\pi \hat{m}_{\ell} v \sqrt{\hat{s} \lambda}-\frac{|{\cal F}|^2\pi \hat{m}_{\ell}(-1+\hat{r}_{K_1}+\hat{s})v\sqrt{\lambda}}{\hat{r}_{K_1}\sqrt{\hat{s}}}
\\ \nnb &+&
\frac{Re[{\cal F}{\cal H}^*]\pi \hat{m}_{\ell}
v(-1+\hat{r}_{K_1}+\hat{s})\sqrt{\hat{s}
\lambda}}{\hat{r}_{K_1}}+\frac{Re[{\cal F}{\cal G}^*]\pi
\hat{m}_{\ell}(-2-2\hat{r}_{K_1}^2+3\hat{s}-\hat{s}^2+\hat{r}_{K_1}(4+\hat{s}))v\sqrt{\lambda}}{\hat{r}_{K_1}\sqrt{\hat{s}}}
\\ \nnb &-&
\frac{|{\cal G}|^2 \pi
\hat{m}_{\ell}(-1+\hat{r}_{K_1})\lambda^{\frac{3}{2}}}{\hat{r}_{K_1}\sqrt{\hat{s}}}+\frac{Re[{\cal
G}{\cal H}^*]\pi \hat{m}_{\ell} v
\lambda^{\frac{3}{2}}\sqrt{\hat{s}}}{\hat{r}_{K_1}}
 \eea

\bea\label{polNN2} \nnb P_{NN}&=&\frac{1}{3}(|{\cal A}|^2-|{\cal
E}|^2)(4 \rl^2-\ss)(3
+3\rr^2-6\ss+3\ss^2-6(1-\ss)-\lambda)+\frac{8Re[{\cal F}{\cal H}^*]
\hat{m}_{\ell}^2\lambda}{\hat{r}_{K_1}}\\ \nnb &+&\frac{8Re[{\cal
G}{\cal H}^*]
\hat{m}_{\ell}^2(-1+\hat{r}_{K_1})\lambda}{\hat{r}_{K_1}}-\frac{4|{\cal
H}|^2 \hat{m}_{\ell}^2 \hat{s} \lambda}{\hat{r}_{K_1}}
\\ \nnb &+&\frac{2Re[{\cal B}{\cal C}^*](-1+\hat{r}_{K_1}+\hat{s})(-6(1+\rr)\ss^2+3\ss^2+\ss(3-6\rr+3\rr^2-\lambda)+4\hat{m}_{\ell}^2\lambda)}{3\ss\hat{r}_{K_1}}
\\ \nnb &+&
\frac{(|{\cal C}|^2-|{\cal
F}|^2)\lambda(-6(1+\hat{r}_{K_1})\ss^2+3\ss^2+\ss(3-6\rr+3\rr^2-\lambda)-4\rl^2\lambda)}{3\ss\hat{r}_{K_1}}
\\ \nnb &+&
\frac{|{\cal
B}|^2(-6(1+\hat{r}_{K_1})\ss^2+3\ss^2+\ss(3+6(-1+8\rl^2)\rr+3\rr^2-\lambda)+4\rl^2\lambda}{3\ss\hat{r}_{K_1}}
\\ \nnb &-&
\frac{|{\cal
G}|^2\lambda(-6(1+\hat{r}_{K_1}+2\rl^2)\ss^2+3\ss^2+\ss(3-6\rr+3\rr^2+24\rl^2(1+\rr)-\lambda)-4\rl^2\lambda)}{3\ss\hat{r}_{K_1}}
\\ \nnb &+&
2\frac{Re[{\cal F}{\cal G}^*]}{3\rr\ss}\Bigg(3(3+4\rl^2+\rr)\ss^3-3\ss^4+(1+4\rl^2-\rr)\ss(3-6\rr+3\rr^2-\lambda)\\
\nnb &-&4\rl^2(-1+\rr)\lambda
\ss^2(-9+6\rr+3\rr^2-24\rl^2(1+\rr)+\lambda)\Bigg)
 \eea

\bea\label{PLN}\nnb P_{LN}&=&\frac{Im[{\cal B}{\cal F}^*]\pi
\hat{m}_{\ell}(-1+\hat{r}_{K_1}+\hat{s})v\sqrt{\lambda}}{\hat{r}_{K_1}\sqrt{\hat{s}}}
\\ \nnb &+&
\frac{Im[{\cal B}{\cal G}^*]\pi \hat{m}_{\ell}
(-1+\hat{r}_{K_1})(-1+\hat{r}_{K_1}+\hat{s})\sqrt{
\lambda}}{\hat{r}_{K_1}\sqrt{\hat{s}}}-\frac{Im[{\cal B}{\cal
H}^*]\pi
\hat{m}_{\ell}(-1+\hat{r}_{K_1}+\hat{s})v\sqrt{\hat{s}\lambda}}{\hat{r}_{K_1}}
\\ \nnb &+&
\frac{Im[{\cal C}{\cal F}^*] \pi
\hat{m}_{\ell}\lambda^{\frac{3}{2}}}{\hat{r}_{K_1}\sqrt{\hat{s}}}+\frac{Im[{\cal
C}{\cal G}^*] \pi
\hat{m}_{\ell}(-1+\hat{r}_{K_1})\lambda^{\frac{3}{2}}}{\hat{r}_{K_1}\sqrt{\hat{s}}}-\frac{Im[{\cal
C}{\cal H}^*]\pi \hat{m}_{\ell}\sqrt{\hat{s}}
\lambda^{\frac{3}{2}}}{\hat{r}_{K_1}}
  \eea

\bea\label{polLN2prim}\nnb P_{NL}&=&-\frac{Im[{\cal B}{\cal F}^*]\pi
\hat{m}_{\ell}(-1+\hat{r}_{K_1}+\hat{s})v\sqrt{\lambda}}{\hat{r}_{K_1}\sqrt{\hat{s}}}
\\ \nnb &-&
\frac{Im[{\cal B}{\cal G}^*]\pi \hat{m}_{\ell}
(-1+\hat{r}_{K_1})(-1+\hat{r}_{K_1}+\hat{s})\sqrt{
\lambda}}{\hat{r}_{K_1}\sqrt{\hat{s}}}+\frac{Im[{\cal B}{\cal
H}^*]\pi
\hat{m}_{\ell}(-1+\hat{r}_{K_1}+\hat{s})v\sqrt{\hat{s}\lambda}}{\hat{r}_{K_1}}
\\ \nnb &-&
\frac{Im[{\cal C}{\cal F}^*] \pi
\hat{m}_{\ell}\lambda^{\frac{3}{2}}}{\hat{r}_{K_1}\sqrt{\hat{s}}}-\frac{Im[{\cal
C}{\cal G}^*] \pi
\hat{m}_{\ell}(-1+\hat{r}_{K_1})\lambda^{\frac{3}{2}}}{\hat{r}_{K_1}\sqrt{\hat{s}}}+\frac{Im[{\cal
C}{\cal H}^*]\pi \hat{m}_{\ell}\sqrt{\hat{s}}
\lambda^{\frac{3}{2}}}{\hat{r}_{K_1}}
  \eea

\bea\label{polNT2}\nnb P_{NT}&=&\frac{4}{3}Im[{\cal A}{\cal
E}^*]\hat{s} v \hat{m}_{\ell} -2(Im[{\cal B}{\cal G}^*]+Im[{\cal
C}{\cal F}^*])
\\ \nnb &&
\frac{v(-1+\hat{r}_{K_1}+\hat{s})(3+3\hat{r}_{K_1}^2-6\hat{s}+3\hat{s}^2-6\hat{r}_{K_1}(1+\hat{s})-\hat{m}_{\ell})}{3\hat{r}_{K_1}}
\\ \nnb &-&
2(Im[{\cal C}{\cal G}^*]\hat{m}_{\ell}+Im[{\cal B}{\cal
F}^*])\frac{v(3+3\hat{r}_{K_1}^2-6\hat{s}+3\hat{s}^2-6\hat{r}_{K_1}(1+\hat{s})-\hat{m}_{\ell})}{3\hat{r}_{K_1}}
\eea

\bea\label{polNT2prim}\nnb P_{TN}&=&-\frac{4}{3}Im[{\cal A}{\cal
E}^*]\hat{s} v \hat{m}_{\ell} -2(Im[{\cal B}{\cal G}^*]+Im[{\cal
C}{\cal F}^*])
\\ \nnb &&
\frac{v(-1+\hat{r}_{K_1}+\hat{s})(3+3\hat{r}_{K_1}^2-6\hat{s}+3\hat{s}^2-6\hat{r}_{K_1}(1+\hat{s})-\hat{m}_{\ell})}{3\hat{r}_{K_1}}
\\ \nnb &-&
2(Im[{\cal C}{\cal G}^*]\hat{m}_{\ell}+Im[{\cal B}{\cal
F}^*])\frac{v(3+3\hat{r}_{K_1}^2-6\hat{s}+3\hat{s}^2-6\hat{r}_{K_1}(1+\hat{s})-\hat{m}_{\ell})}{3\hat{r}_{K_1}}
  \eea

\section{Numerical analysis}
Having the explicit expressions for the physically measurable
quantities, in this section, we will study the dependence of these
quantities on the dileptonic invariant mass($q^2$).
 We will use the parameters given in Tables~\ref{tab:FFinLF} and \ref{input}
in our numerical analysis.
\begin{table}[tbp]
\caption{Input parameters}\label{input}
        \begin{center}
        \begin{tabular}{|l|l|}
        \hline
        \multicolumn{1}{|c|}{Parameter} & \multicolumn{1}{|c|}{Value}     \\
        \hline \hline
         $\alpha_{s}(m_Z)$                   & $0.119$  \\
        $\alpha_{em}$                   & $1/129$\\
        $m_{\konel}$                   & $1.272$ (GeV)\cite{Yao:2006px} \\
        $m_{\koneh}$                   & $1.403$ (GeV) \cite{Yao:2006px}\\
        $m_{\konea} $                  & $1.31$ (GeV) \cite{Yang:2007zt}\\
        $m_{\koneb} $                  & $1.34$ (GeV) \cite{Yang:2007zt}\\
        $m_{b}$                   & $4.8$ (GeV) \\
        $m_{\mu}$                   & $0.106$ (GeV) \\
        $m_{\tau}$                  & $1.780$ (GeV) \\
        \hline
        \end{tabular}
        \end{center}
\label{input}
\end{table}

We present the dependence of the differential single and double
lepton polarization   for the $B \rar K_1 (1272) \ell^+ \ell^-$,
where $\ell=\mu,\, \tau$ decay on $q^2$   as well as its dependence
on $q^2$ due to short distance effects ($\kappa_V\neq0$ case). The
phenomenological factors $\kappa_V$ for the $B \rar K(K^\ast) \ell^+
\ell^-$ decay can be determined from matching the experimental and
theoretical results where they supposed to reproduce correct
branching ratio relation \bea {\cal B} (B \rar J/\psi K(K^\ast) \rar
K(K^\ast) \ell^+ \ell^-) = {\cal B} (B \rar J/\psi K(K^\ast)) {\cal
B} (J/\psi \rar \ell^+ \ell^-)~, \nnb \eea where the right--hand
side is determined from experiments. Using the experimental values
of the branching ratios for the $B \rar V_i K(K^*)$ and $V_i \rar
\ell^+ \ell^-$ decays, for the lowest two $J/\psi$ and $\psi^\prime$
resonances, the factor $\kappa_V$ takes the values:
$\kappa_V=2.7,~\kappa_V=3.51$ (for $K$ meson), and
$\kappa_V=1.65,~\kappa_V=2.36$ (for $K^\ast$ meson). The values of
$\kappa_V$ used for higher resonances are usually the average of the
values obtained for the $J/\psi$ and $\psi^\prime$ resonances. Using
Eq.~(\ref{ratio}) and the results for $\kappa_V$ obtained for
$B\rightarrow K^\ast$ transition\cite{Ali:1999mm}. We find
$\kappa_V=1.75$ for $J/\Psi(1S)$ and $\kappa_V=2.43$ for $\Psi(2S)$,
respectively.

 It is also experimentally useful to consider the averaged values of these
asymmetries. Therefore, we shall calculate the averaged values of
the polarization asymmetries using the averaging procedure defined
as; \bea \la {\cal{P}} \ra = \frac{\ds \int_{4
\hat{m}_\ell^2}^{(1-\sqrt{\hat{r}_{K_1}})^2} {\cal{P}} \frac{d{\cal
B}}{d \hat{s}} d \hat{s}} {\ds \int_{4
\hat{m}_\ell^2}^{(1-\sqrt{\hat{r}_{K_1}})^2} \frac{d{\cal B}}{d
\hat{s}} d \hat{s}}~.\nnb \eea, where ${\cal B}$ is the branching
ratio. The results for averaged value of single and double lepton
polarization asymmetries are presented in table~\ref{average}. Some
of these asymmetries in $B\to K_1(1270)\ell^+\ell^-$ decay(i,e.,
$P_{LL},\,P_{NN}$ and $P_{TT}$ ) are larger than corresponding
asymmetries in $B\to K^*\ell^+\ell^-$ decay presented in
Ref.~\cite{Cornell:2004cp}.

Figs. (1)-(14) show dependence on $q^2$ when considering the
theoretical uncertainties among the formfactors. {\it Note that,
$P_N,\,P_{NL},\, P_{LN},\,P_{NT}$ and $P_{TN}$ for $\mu$ and $\tau$
channels are negligible for all values of $q^2$. Hence, we do not
present their predictions in the figures.}

From these figures, we deduce the following results:
\begin{itemize}
\item{$P_L$ is plotted in Figs. (1) and (2) for muon and tau, respectively.
 It is  decreasing for both of muon and and tau channels. Also, its magnitude  is
much larger for muon channel than tau one. Moreover, there is rather
weak dependency on the theoretical uncertainties among the
formfactors for tau channel.}

\item While {$P_T$ is  decreasing for $q^2\leq 1.2$GeV$^2$ region it is  increasing for  $q^2\geq 1.2$GeV$^2$
region for muon channel(see fig. (3)). Its local minimum at the
point $q^2\leq 1.2$GeV$^2$  is about $-0.15$. $P_T$ is  increasing
in terms of $q^2$ for tau channel(see fig. (4)). Also, $P_T$
vanishes at the end of kinematical region for both muon and tau
channels. }

\item{$P_{LL}$ takes both negative and positive values depending on $q^2$.
 Its zero position occurs at $q^2\simeq 5$GeV$^2$. The measurement of
  the sign of $P_{LL}$ at $q^2\leq 8$GeV$^2$, which is the nonresonance region,
  can be used as a good tool to either check the SM prediction or to search for new physics.
   $P_{LL}$ is quasi uniformly decreasing function of $q^2$ for tau channel.(see figs. (5) and (6)). Moreover, there is rather
weak dependency on the theoretical uncertainties among the
formfactors for tau channel.}

\item{$P_{LT}$ is  decreasing for $q^2\leq 0.8(14.5)$GeV$^2$ region but increasing for  $q^2\geq 0.8(14.5)$GeV$^2$
region for muon(tau) channel (see figs. (7) and (8)). Its local
minimum at the point $q^2\leq 0.8(14.5)$GeV$^2$  is about
$-0.2(0.22)$ for muon(tau) channel, respectively.  Also, $P_{LT}$
vanishes at the end of kinematical region for both muon and tau
channels. Moreover, there is rather weak dependency on the
theoretical uncertainties among the formfactors for tau channel(see
fig. (8)).}

\item{$P_{NN}$ and $P_{TT}$ without resonance contributions are negligible
 at $q^2\geq 8$GeV$^2$ region for muon channel(see figs. (9) and (11)). $P_{TT}$
  takes much larger values in the high $q^2$ region than the low $q^2$ region for tau channel~(see fig. (12)).}

 \item{$P_{TL}$ is  decreasing for $q^2\leq 0.6$GeV$^2$ region but increasing for  $q^2\geq 0.6)$GeV$^2$
region for muon  channel,~(see fig. (13) ). Its local minimum at the
point $q^2\leq 0.6$GeV$^2$  is about $-0.25$. $P_{TL}$ is negligible
for all values of $q^2$ for tau lepton ~(see fig. (14)). Also,
$P_{LT}$ vanishes at the end of kinematical region for both muon and
tau channels. }

\end{itemize}
\begin{table}[tbp]
\caption{Averaged values of single and double lepton
polarizations}\label{average}
        \begin{center}
        \begin{tabular}{|l|l|l|}
        \hline
      \multicolumn{1}{|c|}{$\lla P_{ij}\rra$} & \multicolumn{1}{|c|}{$B \rar K_1(1272)
\mu^+ \mu^-$} & \multicolumn{1}{|c|}{$B \rar K_1(1272)
\tau^+ \tau^-$}     \\
        \hline \hline
      $\lla P_{L}\rra$  & $-0.91\pm0.006$                   & $-0.43\pm0.001$ \\ \hline
       $\lla P_{T}\rra$ &$-0.016\pm0.001$                   &   $-0.05\pm0.004$\\ \hline
       $\lla P_{N}\rra$ &$0.001\pm0.001$                   & $0.01\pm0.001$  \\\hline
      $\lla P_{LL}\rra$  &$-0.34\pm0.0053$                   & $-0.06\pm0.000$\\\hline
       $\lla P_{LN}\rra$ &$-0.001\pm0.000 $                  & $-0.03\pm0.001$ \\\hline
       $\lla P_{NL}\rra$ &$0.001 \pm0.000$                  & $0.03\pm0.001$ \\\hline
       $\lla P_{LT}\rra$ &$-0.06\pm0.003$                   & $-0.17\pm0.000$  \\\hline
      $\lla P_{TL}\rra$  &$-0.03\pm0.003$                   & $-0.01\pm0.000$  \\\hline
       $\lla P_{TT}\rra$ &$0.015\pm0.002$                  & $0.11\pm0.002$  \\\hline
        $\lla P_{NN}\rra$ &$0.01\pm0.004$                  & $-0.17\pm0.001$  \\\hline
         $\lla P_{NT}\rra$ &$-0.006\pm0.001$                  & $0.001\pm0.001$  \\\hline
          $\lla P_{TN}\rra$ &$0.006\pm0.001$                  & $0.001\pm0.001$  \\\hline
        \hline
        \end{tabular}
        \end{center}
\label{input}
\end{table}
Finally, the quantitative estimation about the accessibility to
measure the various physical observables are in order. An
observation of a 3$\sigma$ signal for asymmetry of the order of the
$1\%$ requires about $\sim 10^{12}$ $\bar{B}B$ pairs. The number of
$b \bar{b}$ pairs that are produced at B--factories and LHC are
about $\sim 5\times 10^8$ and $10^{12}$, respectively. As a result,
$q^2$ dependence of the polarization asymmetries shown by
figs.~(1)-(13) as well as averaged values of the same asymmetries
presented in table \ref{average} can be detectable at LHC. Note
that, the ratio of physical observables (for instance, CP ,
foreward--backward and single or double lepton polarization
asymmetries) less suffers from the uncertainty among the formfactors
where large parts of the uncertainties partially cancel out.

 In conclusion, the single and double lepton polarization
asymmetries for exclusive dilepton rare B decays of $B \rar K_1
(1272) \ell^+ \ell^-$ are studied. We have shown that while some
components of lepton polarizations are almost zero, some other
components are sizable to be measured at the future experiments.
Moreover, we show that some of these asymmetries in $B\to
K_1(1270)\ell^+\ell^-$ decay(i,e., $P_{LL},\,P_{NN}$ and $P_{TT}$ )
are larger than corresponding asymmetries in $B\to K^*\ell^+\ell^-$
decay. The study of the magnitude and the size of these physical
observables can be used either to probe the predictions of SM or to
search for new physics effects.

\section{Acknowledgment}
The authors would like to thank T. M. Aliev for his useful
discussions.

\newpage


\newpage

\section*{Figure captions}

{\bf Fig. (1)} The dependence of the $ P_L$ on $q^2$ for
$B\rightarrow K_1(1270) \mu^+\mu^-$ decay, where the colored region
shows
 the variation when theoretical uncertainties among the form factors take into account. \\ \\
{\bf Fig. (2)} The same as in Fig. (1), but for the $\tau$ lepton.\\ \\
{\bf Fig. (3)} The dependence of the $ P_T$ on $q^2$ for
$B\rightarrow K_1(1270) \mu^+\mu^-$ decay, where the colored region
shows the variation when theoretical uncertainties among the form factors take into account.\\ \\
{\bf Fig. (4)} The same as in Fig. (3), but for the $\tau$ lepton.\\ \\
{\bf Fig. (5)} The dependence of the $ P_{LL}$ on $q^2$ for
$B\rightarrow K_1(1270) \mu^+\mu^-$ decay, where the colored region
shows the variation when theoretical uncertainties among the form factors take into account.\\ \\
{\bf Fig. (6)} The same as in Fig. (5), but for the $\tau$ lepton.\\ \\
{\bf Fig. (7)} The dependence of the $ P_{LT}$ on $q^2$ for
$B\rightarrow K_1(1270) \mu^+\mu^-$ decay, where the colored region
shows the variation when theoretical uncertainties among the form factors take into account.\\ \\
{\bf Fig. (8)} The same as in Fig. (7), but for the $\tau$ lepton.\\ \\
{\bf Fig. (9)} The dependence of the $ P_{NN}$ on $q^2$ for
$B\rightarrow K_1(1270) \mu^+\mu^-$ decay, where the colored region
shows the variation when theoretical uncertainties among the form factors take into account.\\ \\
{\bf Fig. (10)} The same as in Fig. (9), but for the $\tau$ lepton.\\ \\
{\bf Fig. (11)} The dependence of the $ P_{TT}$ on $q^2$ for
$B\rightarrow K_1(1270) \mu^+\mu^-$ decay, where the colored region
shows the variation when theoretical uncertainties among the form factors take into account.\\ \\
{\bf Fig. (12)} The same as in Fig. (11), but for the $\tau$ lepton.\\ \\
{\bf Fig. (13)} The dependence of the $ P_{TL}$ on $q^2$ for
$B\rightarrow K_1(1270) \mu^+\mu^-$ decay, where the colored region
shows the variation when theoretical uncertainties among the form factors take into account.\\ \\
{\bf Fig. (14)} The same as in Fig. (13), but for the $\tau$ lepton.\\ \\
\newpage

\begin{figure}
\vskip 1.5 cm
    \includegraphics{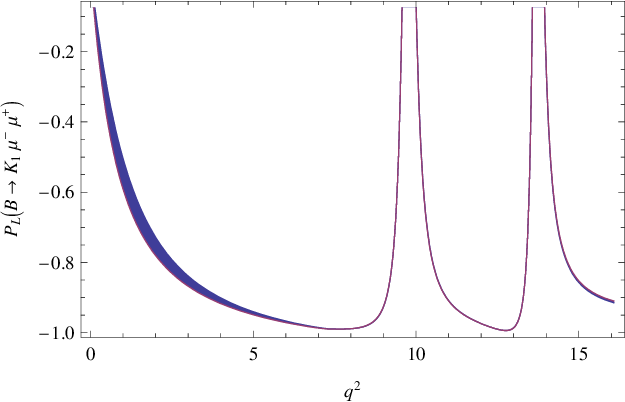}
\vskip 6 cm \caption{}
\end{figure}

\begin{figure}
\vskip 1.5 cm
    \includegraphics{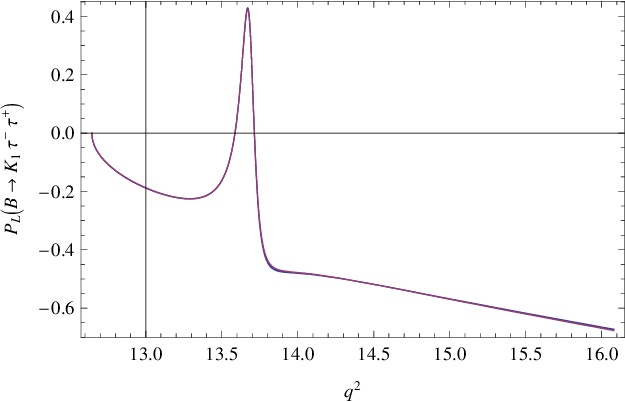}
\vskip 6 cm \caption{}
\end{figure}

\begin{figure}
\vskip 1.5 cm
    \includegraphics{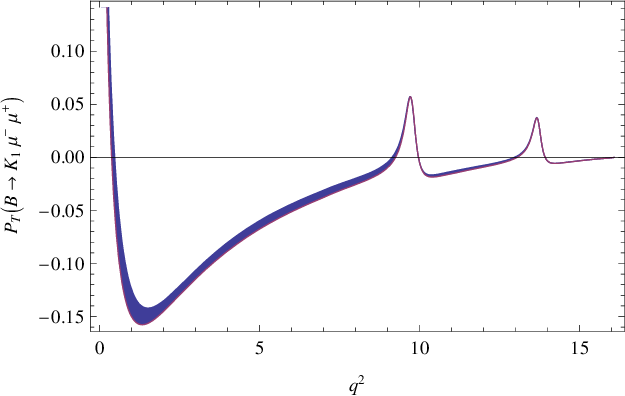}
\vskip 6 cm \caption{}
\end{figure}

\begin{figure}
\vskip 1.5 cm
    \includegraphics{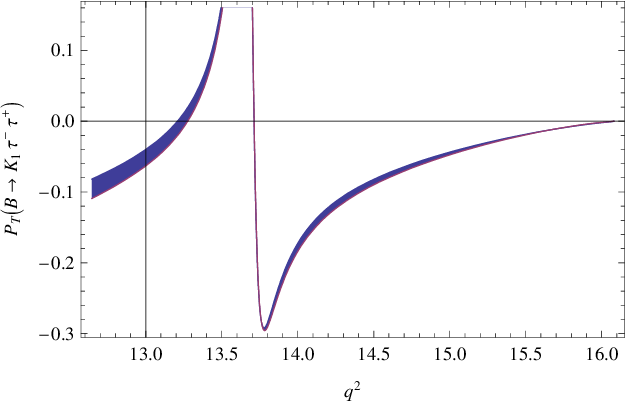}
\vskip 6 cm \caption{}
\end{figure}

\begin{figure}
\vskip 1.5 cm
    \includegraphics{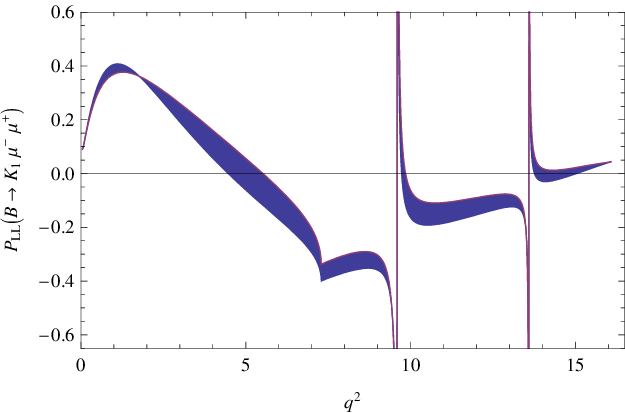}
\vskip 6 cm \caption{}
\end{figure}

\begin{figure}
\vskip 1.5 cm
    \includegraphics{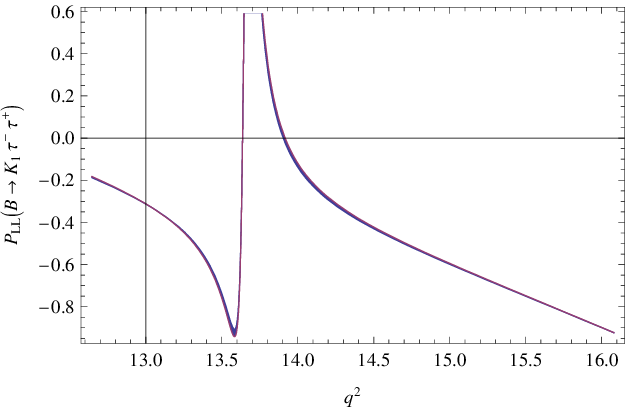}
\vskip 6 cm \caption{}
\end{figure}

\begin{figure}
\vskip 1.5 cm
    \includegraphics{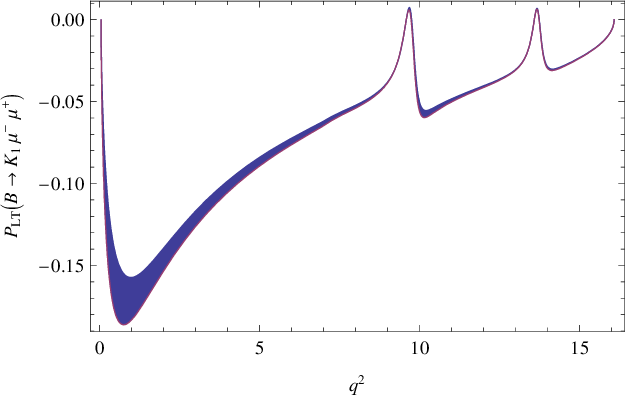}
\vskip 6 cm \caption{}
\end{figure}

\begin{figure}
\vskip 1.5 cm
    \includegraphics{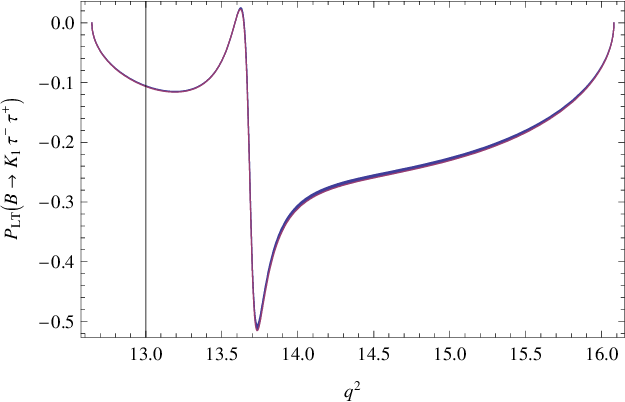}
\vskip 6 cm \caption{}
\end{figure}

\begin{figure}
\vskip 1.5 cm
    \includegraphics{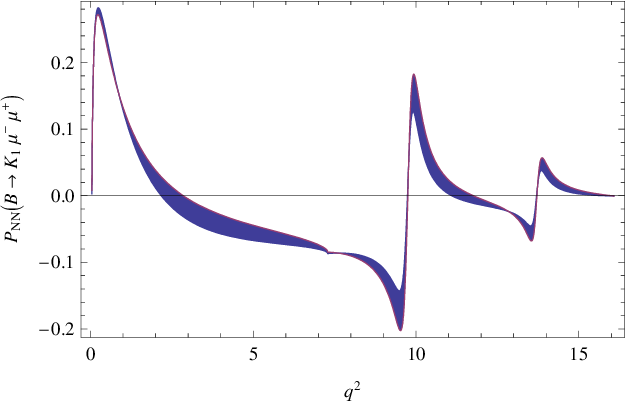}
\vskip 6 cm \caption{}
\end{figure}

\begin{figure}
\vskip 1.5 cm
    \includegraphics{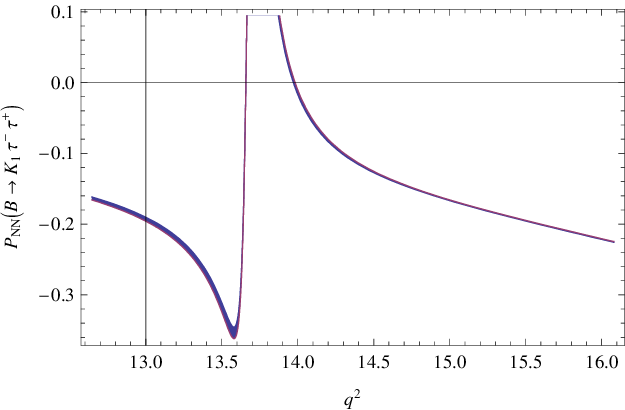}
\vskip 6 cm \caption{}
\end{figure}

\begin{figure}
\vskip 1.5 cm
    \includegraphics{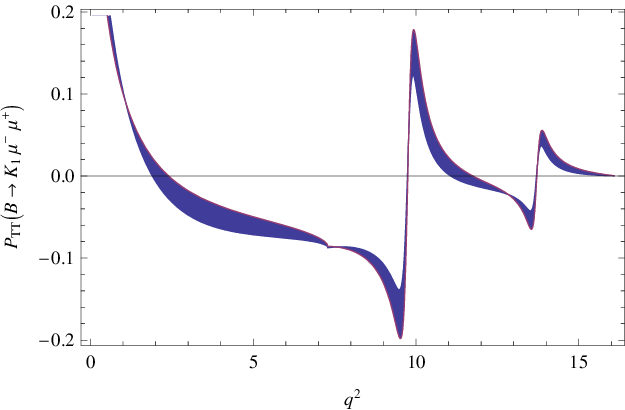}
\vskip 6 cm \caption{}
\end{figure}

\begin{figure}
\vskip 1.5 cm
    \includegraphics{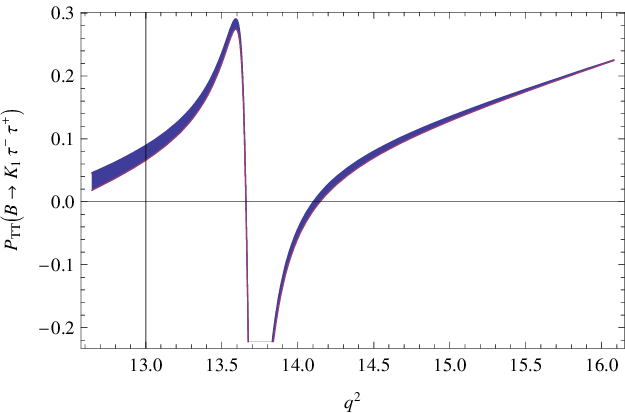}
\vskip 6 cm \caption{}
\end{figure}

\begin{figure}
\vskip 1.5 cm
    \includegraphics{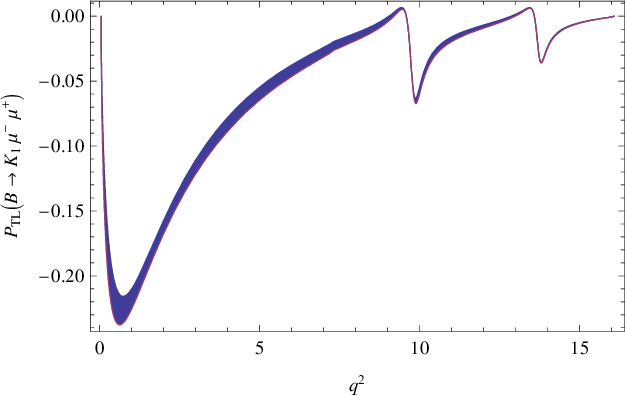}
\vskip 6 cm \caption{}
\end{figure}

\begin{figure}
\vskip 1.5 cm
    \includegraphics{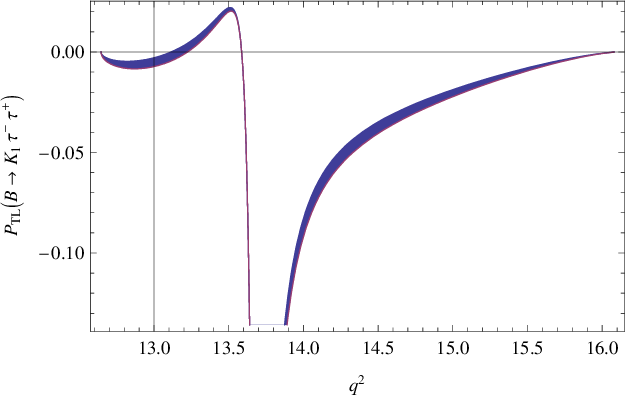}
\vskip 6 cm \caption{}
\end{figure}

\end{document}